# A many-channel FPGA control system

Daniel T. Schussheim and Kurt Gibble[a)]

## AFFILIATIONS

Department of Physics, The Pennsylvania State University, University Park, Pennsylvania 16802, USA

[a)]**Author to whom correspondence should be addressed:** kgibble@psu.edu

## ABSTRACT

We describe a many-channel experiment control system based on a field-programmable gate array (FPGA). The system has 16 bit resolution on 10 analog 100 megasamples-per-second (MS/s) input channels, 14 analog 100 MS/s output channels, 16 slow analog input and output channels, dozens of digital inputs and outputs, and a touchscreen display for experiment control and monitoring. The system can support 10 servo loops with 155 ns latency and MHz bandwidths, in addition to as many as 30 lower bandwidth servos. We demonstrate infinite-impulse-response (IIR) proportional-integral-differential (PID) filters with 30 ns latency by using only bit-shifts and additions. These IIR filters allow timing margin at 100 MS/s and use fewer FPGA resources than straightforward multiplier-based filters, facilitating many servos on a single FPGA. We present several specific applications: Hänsch-Couillaud laser locks with automatic lock acquisition and a slow dither correction of lock offsets, variable duty cycle temperature servos, and the generation of multiple synchronized arbitrary waveforms.

## I. INTRODUCTION

Field programmable gate arrays (FPGA's) are customizable and reconfigurable alternatives to analog electronics to control modern physics experiments. FPGA's often include fast digital logic, digital signal processing (DSP), data transceivers, other hardware elements and reconfigurable interconnections. Combined with high-speed analog-to-digital converters (ADC's) and digital-to-analog converters (DAC's), FPGA's are attractive options for implementing flexible high-speed servos, especially ones that benefit from conditional and dynamic features that are cumbersome to implement with discrete analog components. FPGA's have been widely used for laser and cavity frequency stabilization,[1–10] for phase and frequency metrology[11,12] and laser frequency comb stabilization,[13,14] and for timing pattern generators.[15,16] FPGA servos can provide MHz bandwidths, which are often limited by the latencies of the high-speed ADC's and DAC's that sample at 100 MS/s or higher. A number of high-speed FPGA control systems have been demonstrated that implement one or two servos,[1–3,5–10,13,14] four servos,[17] in addition to a scalable system where an FPGA synchronizes multiple daughter boards, each with its own FPGA that supports two high-speed servos.[4] For slower servos, with sample rates of several MS/s, control systems with as many as 8 servos on a single FPGA have been implemented.[18–20] Systems with many RF inputs, with one or more FPGA's, have been constructed for precise control of RF waveforms for particle accelerators[21–23] and the control of superconducting qubits.[24] A number of these systems use FPGA's integrated into a system-on-chip (SoC),[3–5,7–10,12,14–17,21] which include a processor, facilitating floating point operations, flexible programming, and the implementation of Ethernet and USB communication protocols.

Here, we demonstrate a many-channel FPGA system (MCFS) that uses a single FPGA to implement as many as 10 independent fast servos at 100 megasamples per second (MS/s) (see Fig. 1 and Table I). This MCFS also supports up to 30 slow servo loops, either with analog inputs and outputs or analog inputs and digital outputs. Using a single FPGA facilitates interconnections between multiple servos and with the experiment control, and consumes less power per servo than SoC implementations and systems that use multiple FPGA's. Our system can perform a significant fraction of the tasks in a variety of contemporary experiments, including current atomic physics experiments; we use it to stabilize several lasers and cavities for second-harmonic and doubly-resonant sum-frequency generation,[2,25,26] to laser-cool and trap cadmium.[27–29]

We implement multiple feedback controllers in an FPGA with low-latency digital proportional-integral-differential (PID) gain servos[1,3,4,7,8,10] using fast and efficient infinite-impulse response (IIR) filters.[30] Although some applications, such as high-Q notch filters, require precise filter coefficients, the gain margins of PID servos are often of order 2. Therefore, gain steps and filter coefficients that are $2^n$ often have sufficient precision. Multiplications by coefficients that are $2^n$ are simple and fast bit-shift operations that do not use large multipliers. With one more optional bit-shift and addition for each filter term, our PID gains have a resolution of 25% or better, with coefficients of $2^{-n}(1 + \{-⅛, 0, ¼, ½\})$, ... 0.875, 1, 1.25, 1.5, 1.75, 2, 2.5.... The contributions to the IIR coefficients for PID gains and any pole or zero frequencies are separable. This approach uses a smaller fraction of FPGA resources than multiplier-based filters and can have timing margin at 100 MS/s.

Below we describe our hardware, these bit-shift-addition IIR filters, and several applications that are well-suited for an FPGA control system. One is a servo with automatic locking[1,18] for a build-up cavity for second-harmonic and doubly resonant sum-frequency generation. Here, Hänsch-Couillaud stabilization[31] is enhanced with a slow dither lock to correct lock offsets and their drifts. This lock includes a synthesized dither and a low-resource lock-in amplifier. Another application is a temperature servo for optical cavities and nonlinear crystals that uses a variable-duty-cycle digital output. Finally, we describe synchronized 100 MS/s arbitrary waveform generators that control the laser frequency and intensity for a cadmium magneto-optical trap (MOT) using the narrow 67 kHz wide 326 nm intercombination line.[29] Our MCFS uses a remote touchscreen interface to display current and historical system status and to accept control inputs. Our open-source baseboard design and its associated Verilog software are available online.[32]

TABLE I. Inputs and outputs of the many-channel FPGA system depicted in Fig 1. The fast and slow analog-to-digital converters (ADC's) and digital-to-analog converters (DAC's) have 16 bit resolution. The channels sampled by the slow ADC's can be selected, for example, all channels at 0.125 MS/s or two channels at 1 MS/s.

| Input/Output | # of channels | Sample rate (MS/s) |
|---|---|---|
| Fast ADC | 10 | 100 |
| Fast DAC | 14 | 100 |
| Slow ADC | 16 | 0.125 |
| Slow DAC | 16 | 0.05 |
| Digital I/O | 6+8 | 100 |
| Digital Input | 22 | 2 and 3 |
| Digital Output | 26 | 2 and 3 |

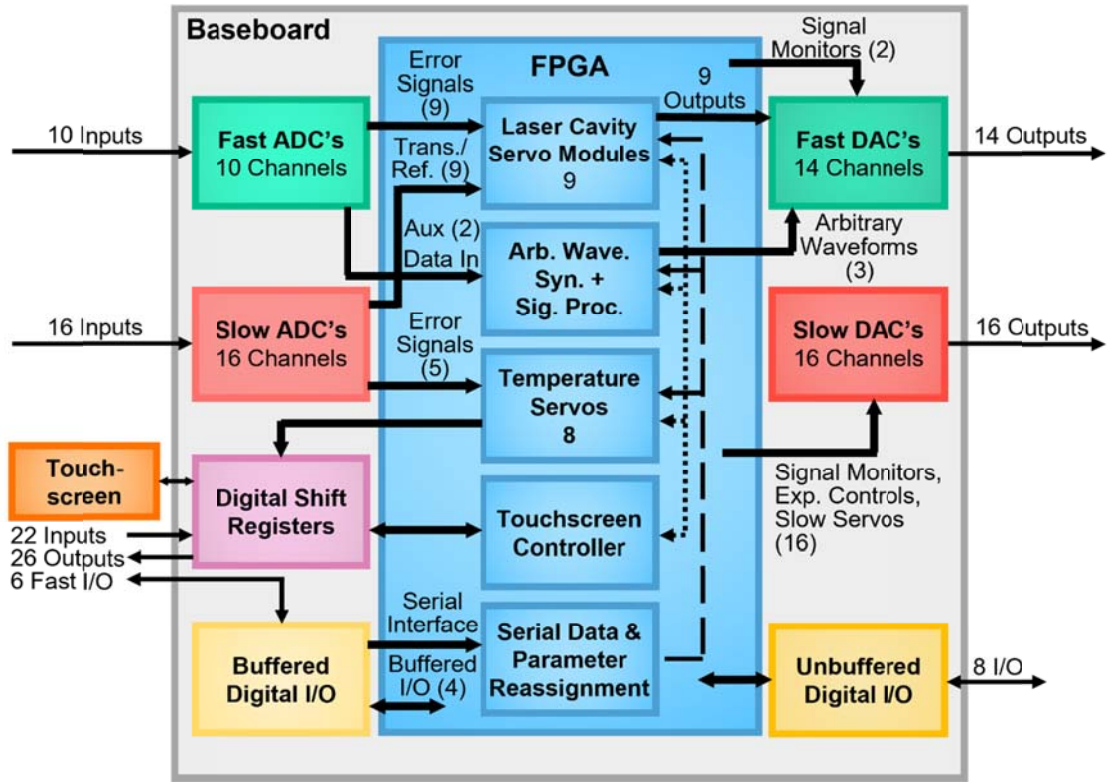

FIG. 1. Schematic of the many-channel FPGA system. An FPGA module and a custom baseboard provide 10 channels of 100 MS/s 16-bit analog-to-digital converter (ADC) inputs and 14 channels of 100 MS/s 16-bit digital-to-analog converter (DAC) outputs. The baseboard also has 16 channels each of multiplexed slow ADC's (125 kS/s) and slow DAC's (50 kS/s), fast digital input/output (I/O) that could interface with additional slow ADC's, and more than 20 digital shift register I/O at 2-3 MS/s, driven by a 50 MHz bus using only 7 FPGA I/O. The FPGA and its software can implement 10 laser/cavity PID servos with automatic lock acquisition and 9 or more variable duty cycle temperature servos, and can be monitored and controlled via the touchscreen display. Our baseline FPGA program has nine laser and cavity servos, eight variable duty cycle temperature servos, an arbitrary waveform synthesizer (Arb. Wave. Syn.) and digital signal processing (Sig. Proc.), a touchscreen display and control interface, and logic to reassign servo and system parameters via a serial data input.

## II. HARDWARE

Our many-channel FPGA system uses a commercial FPGA module[33] that plugs into a baseboard that we developed. The FPGA module has 216 accessible FPGA input/output (I/O), which is sufficient to control the numerous ADC's and DAC's on the baseboard. The FPGA has 25,350 logic slices, each containing 4 look-up tables and 8 flip-flops; 600 DSP slices containing a pre-adder, a 25×18 multiplier, a ternary adder and an accumulator; and 325 36 kb RAM blocks. Other pin-compatible modules with more FPGA resources are available that could accommodate additional software features.

Our 6-layer, 8" × 12" baseboard has 5 two-channel 16-bit 100 MS/s fast ADC's, and 7 two-channel 16-bit 100 MS/s fast DAC's.[34] These converters have 70 ns and 55 ns latency, and use only 10 or 17 FPGA I/O for each 2-channel converter. As in previous FPGA control systems,[1,2,4,6,7,14] the latency of the fast ADC's and fast DAC's are the dominant limitation to the servo bandwidths. In addition to the fast converters, this MCFS has two eight-channel 16-bit slow ADC's and two eight-channel 16-bit slow DAC's (see Table 1). The slow analog channels are useful for lower bandwidth signals and require only 7 and 5 FPGA I/O for the 16 slow ADC and 16 slow DAC channels. The analog inputs and outputs are buffered with operational amplifiers. The fast inputs have 10 MHz bandwidths with a ±4 V range, the fast outputs have 5 MHz bandwidths and a ±18 V range, the slow inputs have 160 kHz bandwidths and a ±10 V range, and the slow outputs have 10 kHz bandwidths and a ±18 V range. The amplifiers and their feedback components are on the opposite side of the board as the ADC's and DAC's, shielding them from digital signals and providing access, e.g., for bandwidth and range modifications, when the baseboard is mounted in its enclosure.

The MCFS also has 6 channels of buffered 100+ MS/s digital I/O, 22 channels of 2 MS/s digital inputs, 26 channels of 2-3 MS/s digital outputs, and 8 channels of unbuffered digital I/O on a FPC connector that could be used for additional slow ADC's. A remote backlit, 3.5" color LCD touchscreen[35] connects to the baseboard via an SPI data bus. The baseboard also has a USB and an Ethernet connector.

The baseboard design reduces digital-analog and analog-analog crosstalk. Ground planes fill much of the unused space on the six layers of the baseboard. Adjacent chips are separated from one another with gaps in the ground planes, especially to guide the return currents of high-speed digital lines. Vias connect the ground planes of each layer to reduce potential differences across ground plane gaps. The ground planes also shield analog signals and power planes from high-speed digital signals. Power is supplied to the baseboard, and in turn to the FPGA module, from a separate circuit board that is fed by a single +15 V input, which drives switching regulators[32] to power the digital electronics and linear regulators for the analog circuits. The switching regulators use frequencies between 0.38 and 1.1 MHz, e.g., to be safely above typical oscillation frequencies of atoms trapped in optical lattices.

We mount the MCFS in an aluminum chassis box, providing heat sinking, radio-frequency shielding, and signal connections for the experiment. Because the FPGA module consumes the highest power of all of the baseboard components, we mount it with a small air gap to an aluminum heat spreader on the side of the box. The FPGA temperature is typically 70 °C with this passive heatsinking, safely below its 100 °C maximum. The ADC's and DAC's temperatures are lower, of order 50 °C, via their heatsinking to the baseboard and convective air currents to the chassis box.

## III. INFINITE-IMPULSE RESPONSE FILTERS

We construct low-latency IIR PID's by summing the outputs of three parallel filters, a $1^{st}$-order proportional (P) filter with a high-frequency roll-off, a $1^{st}$-order integral (I) filter that includes an optional low frequency gain limit, and a $2^{nd}$-order differential (D) filter (Fig. 2). To implement many PID controllers with the MCFS, we use bit-shift-addition IIR filters, which use a smaller fraction of the available FPGA resources than comparable multiplier-based filters. In our design for the configuration of Table II, including real-time adjustability of all parameters, the proportional and integral filters each use a minimum of 1066 (1.1%) FPGA logic slice look-up-tables and the differential uses 1261 (1.2%), for a total of 3.4%. For comparison, multiplier-based filters would use 14 (2.3%) DSP slices each for P and I, and 20 (3.3%) for D, for a total of 8.0%. Filters using bit-shifts have multiplier coefficients of $2^{-n}$, and with an additional single bit-shift-addition, each filter term gives at least 25% resolution, i.e., $2^{-n}(1 + \{-\frac{1}{8}, 0, \frac{1}{4}, \frac{1}{2}\})$. These shift-add filters allow timing margin at 100 MS/s with one clock cycle of latency,[36] whereas our straightforward implementation of multiplier-based IIR filters did not have timing margin.[1,8] In our PID's, the D contribution to the filter output has no additional latency, and we pipeline the addition of the P and I, which delays their contributions by one clock cycle to retain timing margin. Since $1^{st}$-order filters are a subset of $2^{nd}$-order filters, below we first describe a $2^{nd}$-order D filter and then $1^{st}$-order P and I filters, and finally discuss eliminating truncation instabilities of $2^{nd}$-order filters.

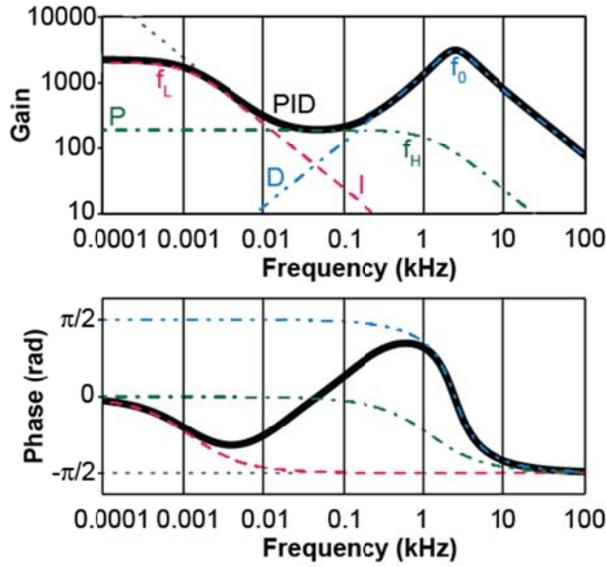

FIG. 2. Gain and phase of a PID transfer function. The PID output (black, solid curve) is the sum of a $1^{st}$-order integral filter (gray dotted curve), including with an optional low-frequency gain cap $I/2\pi f_L$ (red dashed curve), a $1^{st}$-order proportional filter (green dot-dashed curve) with a high-frequency roll-off $f_H$, and a $2^{nd}$-order differential filter (blue, dot-dot-dashed curve) with a high-frequency roll-off $f_0$ and damping $\gamma$.

### A. PID IIR filters

IIR filters are a recursive, discrete-time algorithm that approximates a continuous transfer function with linear combinations of the most recent and prior input(s), and the prior output(s). The output of a general $2^{nd}$-order IIR filter is:

$$y_0 = a_1 y_1 + a_2 y_2 + b_0 x_0 + b_1 x_1 + b_2 x_2,$$

where $y_n$'s are outputs, $x_n$'s are inputs, and $a_n$'s and $b_n$'s are filter coefficients. The subscripts on the $x_n$'s and $y_n$'s indicate previous or current values; $y_0$ is the current output, $y_1$ is the previous output and $y_2$ preceded $y_1$. The filter coefficients, $a_n$ and $b_n$, determine the transfer function,[30] and $a_2 = 0 = b_2$ in first-order filters.

A transfer function for a differential gain $D$ with a high frequency roll-off (see Fig. 2) is:

$$H_D(s) = \frac{D(2\pi f_0)^2 s}{(2\pi f_0)^2 + s(\gamma+s)},$$

where $s = 2\pi i f$, $f_0$ is the roll-off frequency and $\gamma$ is the damping for a filter quality factor $Q = 2\pi f_0/\gamma$. The filter coefficients are:

$$a_1 = 2 - \widetilde{\omega}^2 - \widetilde{\gamma},$$
$$a_2 = -1 + \widetilde{\gamma},$$
$$b_0 = \frac{\widetilde{D}}{2},$$

$$b_1 = 0,$$
$$b_2 = -\frac{\tilde{D}}{2}.$$

Here, $\tilde{\omega} \equiv 2\pi f_0 T/[1 + \gamma T/2 + (2\pi f_0 T)^2/4]^{1/2}$, $\tilde{\gamma} \equiv \gamma T/[1 + \gamma T/2 + (2\pi f_0 T)^2/4]$, and $\tilde{D} \equiv \tilde{\omega}^2 D/T$, where $1/T$ is the filter update rate. The coefficients $a_n$ and $b_n$ separate into gain and frequency terms, $\tilde{D}, \tilde{\omega}^2$ and $\tilde{\gamma}$, and the IIR output becomes:

$$y_0 = y_1 - \tilde{\omega}^2 y_1 + dy - \tilde{\gamma} dy + \frac{\tilde{D}}{2} dx . \qquad (1)$$

Here $dy = y_1 - y_2$ is the difference of the previous two outputs and $dx = x_0 - x_2$ is the difference of the current input and that from two clock periods earlier.[37] We highlight that the differential gain $\tilde{D}$ multiplies only $dx$, and not $y_1$ or $dy$, whereas the filter high-frequency roll-off coefficients $\tilde{\omega}$ and $\tilde{\gamma}$ multiply $y_1$ and $dy$ and not $dx$, beyond $\tilde{\omega}$ scaling the gain. As discussed in more detail in the next section, the desired filter frequencies require a higher precision of $y_0$ than do the gain coefficients, and this naturally allows sub-LSB input servo resolution.

The filter output (1) is the sum of the differential gain contribution, and contributions from the frequency roll-off and filter damping coefficient. Instead of multiplying by the coefficients $a_n$ and $b_n$, the terms for $\tilde{D}, \tilde{\omega}^2$ and $\tilde{\gamma}$ in (1) can be simply implemented with bit-shifts when precisions of factors of 2 are sufficient. For example, a gain $\tilde{D}$ of $2^{-14}$ is a right bit-shift of $dx$ by 14: $dx \gg 14$. For more precise PID contributions, we first optionally add a term with an additional bit-shift before applying the overall shift; $(dx + dx \gg 2) \gg 14$ yields $\tilde{D} = 1.25 \cdot 2^{-14}$. This gives two fractional bits of precision, $2^{-n}(1 + \{-\tfrac{1}{8}, 0, \tfrac{1}{4}, \tfrac{1}{2}\})$, which increases as …0.5, 0.625, 0.75, 0.875, 1, 1.25, 1.5, 1.75, 2…, and similarly for $\tilde{\omega}^2$ and $\tilde{\gamma}$. Along these lines, bit-shifts can be used for coarse scaling, combined with multipliers to retain precision,[6,7] to reduce the required size of the multipliers.

Inverting the above expressions gives $D$, $f_0$ and $\gamma$ in terms of the bit-shifts in (1), $\tilde{D}/2$, $\tilde{\omega}^2$, and $\tilde{\gamma}$. The differential gain is $D = \tilde{D}T/\tilde{\omega}^2$ with a high frequency roll-off $f_0 = \tilde{\omega}/2\pi T/[1 - \tilde{\gamma}/2 - \tilde{\omega}^2/4]^{1/2}$, and damping $\gamma = \tilde{\gamma}/T/(1 - \tilde{\gamma}/2 - \tilde{\omega}^2/4)$, where $\tilde{D}$, $\tilde{\omega}^2$, $\tilde{\gamma} = 2^{-n}(1 + \{-\tfrac{1}{8}, 0, \tfrac{1}{4}, \tfrac{1}{2}\})$. We note that $f_0$ and $\gamma$ become nonlinear in $\tilde{\gamma}$ and $\tilde{\omega}$ for large $\tilde{\gamma}$ and $\tilde{\omega}$. To have timing margin at 100 MS/s, we use two fractional bits of precision for $\tilde{D}$ and $2^{-n}$ precision for $\tilde{\omega}^2$ and $\tilde{\gamma}$, which gives $2^{-1/2}$, 1, $2^{1/2}$, 2, … resolution for $\tilde{\omega}$. Although the implementation timing report may not show timing margin for differential filters that have two fractional bits of precision for $\tilde{\omega}^2$ and $\tilde{\gamma}$, we nonetheless observed reliable operation at 100 MS/s. Further, if the differential gain $\tilde{D}$ remains adjustable and the high-frequency roll-off and damping are fixed, $\tilde{D}, \tilde{\omega}^2$ and $\tilde{\gamma}$ can all have two fractional bits of precision with timing margin at 100 MS/s. For the update rates of our temperature servos, this filter has timing margin with adjustable 25% precision on all terms.

We similarly follow the above steps for the D filter for first-order P and I filters, with transfer functions:

$$H_P(s) = \frac{P}{1 + s/2\pi f_H} \text{ and}$$

$$H_I(s) = \frac{I}{2\pi f_L + s}.$$

Here, $P$ is the proportional gain, $f_H$ is a high-frequency roll-off, $I$ is the integral gain, which can include a low-frequency integral gain limit of $I/2\pi f_L$. These P and I filters have functionally identical coefficients:

$$a_1 = 1 - \tilde{\omega}_{H/L}$$

$$b_0 = b_1 = \frac{\tilde{G}}{2}$$

where, $\tilde{\omega}_{H/L} \equiv 2\pi f_{H/L} T/(1 + 2\pi f_{H/L} T/2)$ and $\tilde{G} \equiv \tilde{\omega}_H P$ or $IT$ for the P and I filters. The filter output can then be written as:

$$y_0 = y_1 - \tilde{\omega}_{H/L} y_1 + \frac{\tilde{G}}{2} sx \tag{2}$$

where $sx = x_0 + x_1$. We implement (2) with bit-shifts and additions, as for the D filter above. Inverting the expressions gives $P = \tilde{G}/\tilde{\omega}_H$, $I = \tilde{G}/T$ and roll-off frequencies $f_{H/L} = \tilde{\omega}_{H/L}/2\pi T/(1 - \tilde{\omega}_{H/L}/2)$, where $\tilde{\omega}_{H/L} = 2^{-n}(1 + \{-⅛, 0, ¼, ½\})$ and $f_{H/L}$ are again nonlinear in $\tilde{\omega}_{H/L}$. These filters can have timing margin at 100 MS/s with adjustable parameters that have two fractional bits of precision.

TABLE II. PID gains and frequencies for 100 MS/s filters. These values are for I and P filters with 16+9+32 bits and a D filter with 16+9+16 bits, as discussed in the text. The PID gains and $f_{L/H}$ can be zero, and the minimum nonzero values are given. The minimum $P$ gain depends on $f_H$, and the table shows the minimum nonzero values of $P$ at the minimum and maximum $f_H$. Similarly, $D$ depends on $f_0$ and $\gamma$, and the minimum values of $D$ are shown for the minimum and maximum $f_0$ for $Q \approx 1$. Normally, the maximum gains are not a limitation when servos have LSB resolution and use a high-frequency filter clock.

|  | Minimum Gain | Frequency Response | |
|---|---|---|---|
| I | 0.18 rad s$^{-1}$ | $f_L$ | 0, 7.2 µHz |
|  |  |  | 32 MHz |
| P | 4032 | $f_H$ | (0,) 7.2 µHz |
|  | 1.8×10$^{-9}$ |  | 32 MHz |
| D @ Q ≈ 1 | 4.0×10$^{-5}$ rad$^{-1}$ s |  | 2.7 kHz |
|  | 1.2×10$^{-12}$ rad$^{-1}$ s | $f_0$ | 32 MHz |
|  |  | $\gamma$ | 3 s$^{-1}$ – 2×10$^8$ s$^{-1}$ |

    Our minimum PID latency is τ = 155 ns: 125 ns from the fast ADC and DAC conversions, 10 ns for the fast ADC firmware, 10 ns from the fast DAC firmware, and 1 clock cycle, 10 ns, from the PID filters. If the servo is stable with π/2 phase margin, the maximum servo bandwidth is then 1/4 τ = 1.6 MHz.

## B. Fractional bits, filter stability, and rounding

IIR filters that sample much faster than the servo bandwidth produce less aliasing and a more linear servo response. A straightforward implementation of the above PID filters then requires using words in the filter that are longer than our 16-bit input and output word to allow low-frequency integral gain limits and high-frequency roll-offs that are far below the sampling rate. The gain and frequency ranges for internal words with 16 + 9 + 32 = 57 bits for our P and I filters, and 16 + 9 + 16 = 41 bits for the D filter, are given in Table II for 100 MS/s. Here, the 16 most significant bits correspond to the inputs and outputs from the ADC's and DAC's. The inputs to the PID filters have 9 fractional bits of precision, allowing sub-LSB corrections to the PID inputs. Finally, to enable low filter frequencies, the PID filters have an additional 32 or 16 internal fractional bits. Here, the 9 servo fractional bits and the 32 or 16 internal fractional bits both extend the lower range of filter frequencies, whereas only the 32 or 16 internal fractional bits yield lower gains. Therefore, increasing an unnecessarily small minimum filter gain can allow higher input resolution for a given filter internal word size. With the ranges in Table I, our PID's have timing margin at 100 MS/s. For comparison, a straightforwardly implemented multiplier-based filter with the same parameter ranges and $2^{-n}(1 + \{-⅛, 0, ¼, ½\})$ precision requires 56-bit filter coefficients, which are long enough that straightforwardly implemented filters do not have timing margin at 100 MS/s.

Second and higher-order filters can be unstable as errors accumulate due to the truncation of least-significant bits. For example, the term $-\tilde{\gamma}dy$ in (1) of the D filter yields a slow decay of $dy$. This decay ceases when $-\tilde{\gamma}dy$ is smaller than the least-significant bit (LSB) of the 41-bit internal filter word. The filter thus would continue to add $dy$ in (1) to make the new output $y_0$, which will normally cause $y_0$ to grow until it overflows. To avoid this accumulation error, we assign $\tilde{\gamma}dy$ to be ±1 LSB of the 41-bit word when $0 < \pm\tilde{\gamma}dy < 1$. Finally, we round numbers before truncating the LSB's when applying right bit-shifts; we first add $2^{s-1}$ before dividing by $2^s$, a right bit-shift of $s$.[38]

## IV. SELECTED MCFS APPLICATIONS

### A. Hänsch-Couillaud stabilization with a slow dither lock correction

We use Hänsch-Couillaud (HC) cavity locks to stabilize several laser frequencies and optical cavity lengths in our laser system. HC locks have low loss and high bandwidth, but can suffer from slow lock offset drifts, for example due to temperature dependent birefringences. To correct lock offsets and their drifts, we augment HC locks with slow dither locks to the peak transmission, minimum reflection, or peak sum-frequency generation (SFG) output of a cavity.[39] Dither locks of lasers and optical cavities add frequency modulation at the dither frequency, as well as intensity modulation at twice the modulation frequency that is proportional to the square of a small dither amplitude. Here, because the cavity is primarily locked by the higher bandwidth HC lock, only a small dither amplitude is required to correct lock offsets, and thus it produces a very small intensity modulation. In our locks, the amplitude of the dither is well below the root-mean-square (RMS) noise level of the closed-loop error signal within a typical servo bandwidth of 40 kHz, and even well below the noise in a 1 kHz bandwidth for a dither frequency of order 1 kHz. We normally use dither lock servo bandwidths of order 20 mHz and the MCFS further includes logic to inhibit dithers, for example, when lasers are pulsed for laser-induced fluorescence detection.

We implement laser and cavity servos with automatic lock acquisition[1–3,5,7,8,10,18] and a slow dither lock correction. To acquire lock, a servo output is scanned until a cavity transmission, reflection, or SFG output passes a threshold, at which point a PID filter is enabled. A feature we find very helpful is displaying each servo's lock status with one of three colors, indicating that the servo is unlocked, locked for longer than 5 seconds, or recently locked, having been unlocked within the last 5 seconds. To correct lock offsets, a synthesized dither is added to the Fast Error Signal in Fig. 3(a), modulating the transmission, reflection, or SFG output, which is then demodulated by a lock-in amplifier to form the slow error signal with high long-term stability. This slow error signal is integrated to correct any offset of the Fast Error Signal. The dither is synthesized from a simple stepped waveform [dotted green curve in Fig. 3(b)] that has no $3^{rd}$ harmonic and reduced $5^{th}$ and $7^{th}$ harmonics. Integrating it twice (dashed blue and solid black) reduces the higher odd harmonics to form a nearly sinusoidal dither, ranging from 93 µHz to 1.67 MHz. We use a simple demodulation waveform (red dashed) that also contains no $3^{rd}$ harmonic. Similar integrations demodulate the quadrature $1^{st}$ harmonic and the in-phase and quadrature $2^{nd}$ and $3^{rd}$ harmonics. We note that incorporating bit-shift-addition operations, or a multiplier, instead of this simple 3 level demodulation would slightly increase the demodulated signal-to-noise and further reduce the sensitivity to $5^{th}$ and higher odd harmonics.

The cavity lock for our SFG of 361 nm light, from 1083 nm and its second harmonic, 542 nm, is another example of the flexibility that an FPGA affords. We use the above HC lock with its slow dither correction to lock a doubly resonant enhancement cavity to the 542 nm light. Because the 542 nm is the $2^{nd}$ harmonic of the 1083 nm light, the locked enhancement cavity largely tracks the frequency of the 1083 nm input and only a slow correction of its frequency is required, provided by an acousto-optic modulator driven by a voltage-controlled oscillator (VCO). We therefore use a dither lock to lock the 1083 nm light to the enhancement cavity. However, the slow dither lock of the 542 nm lock can interfere with the 1083 nm dither lock. To avoid this, we configure the FPGA to alternately dither the 542 nm error signal or the 1083 nm frequency, while inhibiting the other. Here, we use the intensity of the 361 nm SFG light to enable the PID's and for both dither locks, thereby maximizing the SFG output.[39] As for other locks, we inhibit both dithers for laser-induced fluorescence detection.

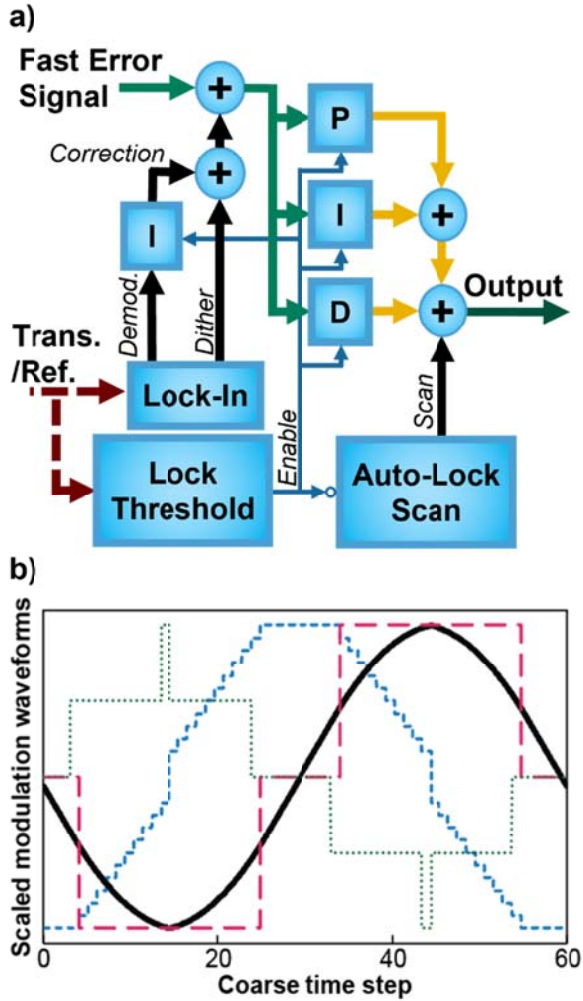

FIG. 3. (a) Schematic of a cavity lock with a correction from a slow dither lock. The cavity frequency is scanned and, when the cavity transmission (Trans.) or reflection (Ref.) passes a threshold, the PID filter is enabled. A dither is added to the Fast Error Signal and the resulting modulation of the transmission or reflection is demodulated (Demod.). This is then integrated to give the Correction of the offset of the Fast Error Signal. (b) Modulation waveforms. The dither is synthesized from the dotted green curve, by integrating it twice (dashed blue and solid black), producing a dither with no $3^{rd}$ and reduced higher odd harmonics. Adjusting the coarse time steps provides dither frequencies from 93 µHz to 1.67 MHz. The demodulation waveform (red dashed) also contains no $3^{rd}$ harmonic.

### B. Variable duty cycle temperature servo

We implement several servos using the slow ADC's and digital outputs to control the temperatures of non-linear crystals, a reference cavity and a heated Cd oven. Such systems often have thermal response times of order 0.1 s to 100 s and variable duty cycle (VDC) servos can easily be implemented with the FPGA. As compared to linear current regulation, pulse width modulation uses less power, with negligible added temperature noise for pulse periods much shorter than the system's response time. With a single FPGA controlling multiple servos, it is straightforward to synchronize the delays of the pulses of multiple servos to provide load diversity for a single power source.

Figure 4 depicts a VDC temperature PID servo that produces a constant frequency output with an adjustable duty cycle. As discussed above, fixing filter coefficients, such as filter roll-off frequencies, $f_L, f_H, f_0$ and damping $\gamma$, yields more timing margin and significantly reduces the required resources. Often, the frequency response only changes significantly when the plant being controlled is substantially modified so adjustable $f_H, f_0$ and $\gamma$ are not needed. Further, the frequency response of the plant determines the ratio of the proportional and integral gain, and the ratio of the differential and proportional. We therefore include a multiplier after the sum of the PID gains in Fig. 4 that allows the overall gain to be adjusted even when $f_L, f_H, f_0$ and $\gamma$, as well as the *P*, *I* and *D* gains, are not adjustable.[40] This saves significant resources and has timing margin for low filter clock frequencies. A seven-bit (signed) multiplier allows the gains to be adjusted in steps of 1/16, from ¼ to greater than 2 with greater than 25% precision. We use a 125 kHz clock for our temperature servos, which naturally gives lower ranges for the filter frequencies and smaller gains, and matches the sample rate of the slow ADC's when all channels are sampled sequentially.

Using shift-register outputs to switch heater currents uses only a few high-speed FPGA outputs to control multiple temperature servos. However, with a typical 1 kHz VDC frequency, our 2 MS/s shift-register update rate corresponds to a duty cycle resolution of 0.05%. We increase this resolution by a factor of 16, when averaged over 16 cycles of 1 kHz, by successively adding {0, 15, 1, 13, 3, 11, 5, 9, 7, 8, 6, 10, 4, 12, 2, 14}/16 to the PID output, before the output is rounded to an integer number of 2 MS/s samples. This sequence minimizes the noise by modulating the LSB slowly, and the most-significant fractional bit on every 1 kHz cycle. As an example, consider a PID output of 82.664%, corresponding to 1,653.28 samples at 2 MS/s during each 1 kHz VDC cycle. Successively adding the above sequence over 16 cycles of 1 kHz truncates the PID output 12 times to 1,653 cycles and rounds four times to 1,654, for an average of 1,653.25 cycles.

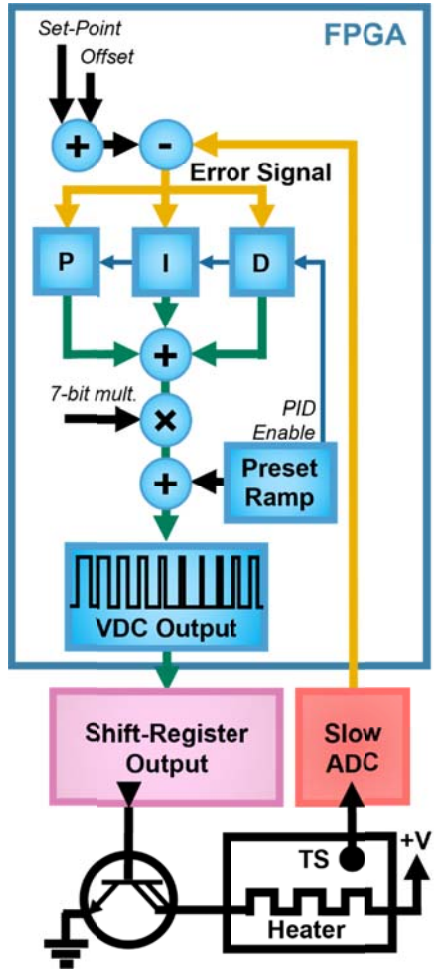

FIG. 4: Variable duty cycle temperature servo. A slow ADC reading a temperature sensor (TS), relative to an optional setpoint and offset, produces an error signal for a PID servo. The servo output is added to a preset to drive a variable duty cycle digital shift-register output, which pulses current through a heater at a typical rate of 1 kHz. To avoid thermal shocks, before the PID is enabled, the preset increases slowly, on a timescale of order minutes.

### C. Arbitrary waveform generation

The MCFS's 14 channels of 100 MS/s DAC's can generate multiple synchronized arbitrary waveforms with 10 ns resolution. Figure 5 shows three synchronized waveforms generated by a counter-driven state machine. This approach allows longer high-sampling-rate waveforms than possible with memory-based AWG's. We use the AWG to control the laser frequency (blue-solid) and intensity (green-dashed) and trigger a magnetic field gradient driver to trap neutral cadmium using its 326 nm, 67 kHz wide intercombination transition.[29] Note that the MCFS allows the frequency modulation during the loading stage of the magneto-optical trap (MOT) to always end (and begin) without an abrupt frequency step. We use the two-level trigger (magenta-dotted) to synchronize the reversal of the MOT magnetic field gradient for background subtraction. A touchscreen display button conveniently allows changes between waveforms for several configurations of the experiment.

To sensitively detect the fluorescence of trapped atoms, we implement a gated integrator with background subtraction. In Fig. 5, during the "+" detection phase, with no laser FM, the fluorescence signal is integrated for a time $\Delta t_{int}$ = 16.6716 ms, approximately one 60 Hz cycle. In the subsequent $\Delta t_{int}$ interval, the laser frequency is tuned to the blue of the transition to expel the cold atoms from the trap and then the background is integrated in the next interval of $\Delta t_{int}$, "-", and subtracted from the gated integration of the fluorescence. This difference of gated integrations is stored in Block RAM and can be read from the FPGA. Additionally, the MOT magnetic field gradient is reversed after each trapping and detection sequence and the difference of gated integrations from one cycle to the next are subtracted and stored, representing the difference in fluorescence for a trapping or anti-trapping MOT magnetic field gradient. These gated integrations with background subtraction and the difference of successive integrations are also connected to fast DAC's and can be displayed on an oscilloscope in real-time.

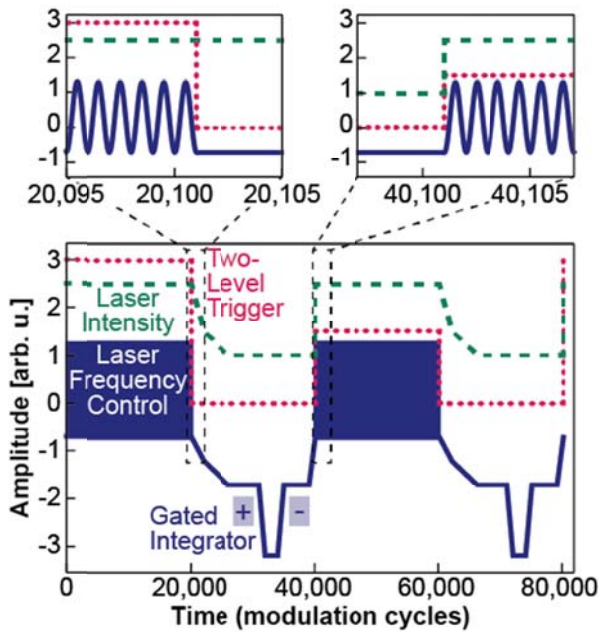

FIG 5. Three synchronized 100 MS/s arbitrary waveforms, adjustable in real-time, to control a laser frequency (blue-solid) and intensity (green-dashed), and trigger MOT field gradients (magenta-dotted) to laser-cool neutral cadmium. The laser light is frequency modulated with an acousto-optic modulator at 50.5 kHz for approximately 400 ms during the MOT loading phase, and then shifted to higher frequency (lower voltage) during a clearing pulse. We use a state machine architecture to produce synchronized long arbitrary waveforms.

## V. Conclusion

We demonstrate a many-channel system using a single FPGA to control a large number of experimental sub-systems, including high-speed PID laser and cavity locks, temperature controllers, synchronized arbitrary waveform generation, and the experiment configuration with a remote touchscreen display. We also demonstrate an enhanced Hänsch-Couillaud cavity lock, where offsets are corrected with a very small amplitude dither-lock, as well as variable-duty-cycle temperature servos. Implementing PID IIR filters with bit-shifts and additions allows real-time adjustment of servo gains with 25% precision, with timing margin at 100 MS/s, and uses fewer FPGA resources than multiplier-based filters.

A number of options can provide more available logic, including transferring more operations to the many available DSP slices in our design and using pin-compatible FPGA modules with significantly more resources. Hard-coding PID roll-off frequencies,[40] $f_L, f_H, f_0$ and $\gamma$, with 25% precision uses half as many look-up tables while retaining real-time adjustment of the PID gains and thereby the zeroes of the PID transfer function. Restricting the ranges of gains, fixing the relative PID gains and allowing only an overall gain adjustment, or less precision of the gain or high frequency roll-offs, all save additional FPGA resources. Our default configuration, with arbitrary waveform generation and DSP, has nine cavity servos and two temperature servos that are fully adjustable, and six temperature servos with fixed PID parameters and adjustable overall gains. Additionally, the operations of PID's that update at less than 100 MS/s, such as the temperature servos, could be pipelined so that a single PID filter sequentially implements multiple temperature servos. Finally, the proportional, integral and differential filters can be pipelined to use the same logic slices[1] and the internal word lengths of the filters can be shortened if the ranges in Table II are not required. Thus, as many as 10 fast servos and 30 slow servos, after adding a daughter board with 24 additional slow ADC channels, could be implemented on a single FPGA with this control system. The open-source software and hardware files for this system are available[32] to facilitate extending and customizing this many-channel FPGA system for a variety of applications.

## ACKNOWLEDGEMENTS

We gratefully acknowledge many suggestions from Avrum Warshawsky, contributions of Lam Tran, helpful conversations with Marco Pomponio, and financial support from the National Science Foundation.

## DATA AVAILABILITY

The supporting files for this open-source many-channel FPGA system are available at https://github.com/GibbleLab/FPGA.

## APPENDIX: Input and Output Noise

The analog input and output noise of the MCFS are shown in Fig. 6 and are primarily set by the ADC and DAC noise levels. To measure the noise of the ADC's in Figs. 6(a-c), the inputs were terminated and their outputs were read by FPGA debugging probes. In Fig. 6(d-f) the DAC's were programmed to output 0 and their noise was measured with a fast ADC. The measurement noise level of the fast ADC's in Figs. 6(d-f) is 4/18 of that in Fig. 6(a,b), after accounting for the 4 V input and 18 V output ranges. The average measured RMS noise levels are 3.7 LSB for the fast ADC's, 1.13 LSB for the fast DAC's in a 10 MHz bandwidth, 0.48 LSB for the slow ADC's, and 0.16 LSB for the slow DAC's in a 200 kHz bandwidth. The coherent peak in Figs. 6 at 380 kHz is from a -20 V switching supply on our power supply board. Its RMS amplitude in Fig. 6b) is 0.028 LSB, and an average of 0.015 LSB for the 10 fast ADC's, 0.050 LSB for the 14 fast DAC's, and 0.017 LSB for the 16 slow DAC's. The frequencies of the other switching supplies on our power supply board are greater than 600 kHz and below the noise levels in Fig. 6. The largest coherent peaks in Fig. 6f) are from glitches at multiples of the update rate of the slow DAC's, here at 50 kS/s. To reduce the glitch amplitude, the MCFS baseboard has 5th-order low-pass filters on the slow DAC outputs that strongly attenuate frequencies above 300 kHz, with less than π/4 phase lag at frequencies below 10 kHz. This yields an average glitch amplitude of 0.36 LSB from an average glitch impulse of –3.0 LSB·µs. To decrease crosstalk between the fast ADC and DAC channels, the MCFS baseboard has slots in the multiple ground and power planes and between adjacent channels and converters. We measure –70 dBc crosstalk for a 1 MHz full scale (±4 V) input of a fast ADC on the other channel of the same ADC, less than –80 dBc on channels of the other fast ADC's, and the attenuation is higher at lower frequencies. Finally, the distribution of the bipolar offset errors of the 14 fast DAC outputs have a standard deviation of 1.9 mV and a mean of 1.2 mV. An appropriate DAC channel can thus be selected to reduce the bipolar error.

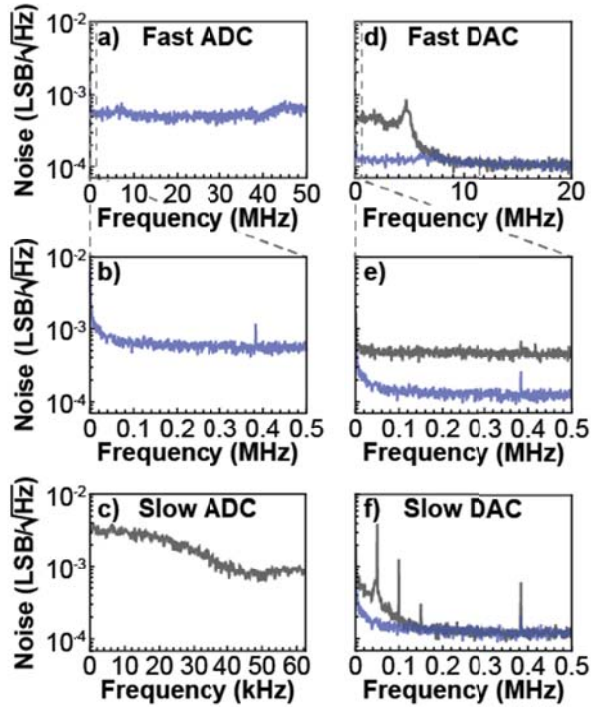

FIG. 6. Input and output noise spectral densities. The fast ADC [blue in a) & b)] is used to measure the noise of the fast and slow DAC's (d-f), and its noise floor is shown in (d-f), shifted by the 4 V/18 V ratio of the ranges of the inputs and outputs. The 380 kHz peak from a switching regulator has an RMS amplitude less than 0.034 LSB on all ADC's and DAC's. The peaks in f) are at multiples of the 50 kS/s sampling frequency of the slow DAC's, due to intrinsic glitches of the slow DAC's, and correspond to an average RMS amplitude of 0.12 LSB. All data were sampled at 100 MS/s with a fast ADC, except for c), which was sampled at the maximum 125 kS/s of the slow ADC's. The data for e) and f) were additionally averaged with a 100-sample window and down-sampled at 1 MS/s.

[37] We observed dx being synthesized as (x0>>>bs)-(x2>>>bs), so retaining x0>>>bs for subsequent calculations, and similarly for dy, increases the timing margin.

[38] A single fast PID servo in our design has timing margin with $2^{-n}(1 + \{-\frac{1}{8}, 0, \frac{1}{4}, \frac{1}{2}\})$ resolution for the proportional and integral gains and rolloff frequencies. While our normal full design with nine such fast PID's does not have timing margin, we have not observed glitches or other errors using it. Restricting the proportional and integral rolloff frequencies to $2^{-n}$ resolution, making the proportional and integral rolloff frequencies non-adjustable with $2^{-n}(1 + \{-\frac{1}{8}, 0, \frac{1}{4}, \frac{1}{2}\})$ resolution, or omitting the rounding of the $\widetilde{\omega}_{H/L}$ term in (2) provide timing margin at 100 MS/s in our full design.

[39] E. Mimoun, L. De Sarlo, J.-J. Zondy, J. Dalibard, and F. Gerbier, "Sum-frequency generation of 589 nm light with near-unit efficiency," Opt. Express **16**, 18684 (2008).

[40] The adjustability of parameters can be easily eliminated by simply removing the assignment(s) of the updated value(s). After the change, resynthesizing and reimplementing the code is required.